# Let Community Rules Be Reflected in Online Content Moderation

Wangjiaxuan Xin

University of North Carolina at Charlotte

wxin@charlotte.edu

Zhe Fu

University of North Carolina at Charlotte

zfu2@charlotte.edu

Kanlun Wang

Fairfield University

kanlun.wang@fairfield.edu

Lina Zhou

University of North Carolina at Charlotte

lzhou8@charlotte.edu

**Abstract**

*Content moderation is a widely used strategy to prevent the dissemination of irregular information on social media platforms. Despite extensive research on developing automated models to support decision-making in content moderation, there remains a notable scarcity of studies that integrate the rules of online communities into content moderation. This study addresses this gap by proposing a community rule-based content moderation framework that directly integrates community rules into the moderation of user-generated content. Our experiment results with datasets collected from two domains demonstrate the superior performance of models based on the framework to baseline models across all evaluation metrics. In particular, incorporating community rules substantially enhances model performance in content moderation. The findings of this research have significant research and practical implications for improving the effectiveness and generalizability of content moderation models in online communities.*

## 1. Introduction

Social media platforms have become prevalent channels for users to collaborate and share knowledge and experiences across diverse online communities. Nevertheless, it also opens a door to the propagation of irregular content, including unsubstantiated or false information. Without adequate moderation, such dissemination on social media poses risks to the integrity and trustworthiness of the online community (Allcott et al., 2019). Most social media platforms have adopted a governance mechanism, which aims to "structure participation in a community to facilitate cooperation and prevent abuse" (Grimmelmann, 2015). Content moderation is a typical intervention strategy for regulating online communities on social media platforms, to ensure that user-generated content complies with the platforms' policies and community standards (Gillespie, 2020).

With the advancement of AI technologies and the increasing workload associated with online moderation (Batrinca & Treleaven, 2015), online platforms are increasingly adopting machine learning and/or deep learning-based techniques to automate content moderation, particularly to address its scalability issue (Gorwa et al., 2020). Despite the fact that the automated approaches to online content moderation have become a significant research focus (Gongane et al., 2022), particularly with deep learning-based frameworks demonstrating success in content moderation (e.g., Founta et al., 2019; Iwendi et al., 2023; K. Wang et al., 2023), most of these studies design deep neural networks as black-boxes (Bunde, 2021).

Community rules are a distinctive characteristic of an online community ecosystem (Fiesler et al., 2018). These rules can be categorized into three levels: macro community norms, which consist of universal rules applicable to all online communities; meso norms, which are guidelines enforced for a group of online communities; and community-created micro norms, which are specialized to individual and unique communities (Chandrasekharan et al., 2018). These rules serve as a crucial tool for online moderators to manage online content and maintain the community ecosystem. By complying with these rules and norms, community moderators govern and regulate user behaviors, including posts and comments, to ensure a healthy and courteous social platform environment (Jhaver et al., 2019). This study is focused on community-created micro norms.

There are very limited studies on community rule-based content moderation. Cheng et al., (2015) utilized random forest to predict antisocial behaviors in particular communities (e.g., Breitbart.com in political news and IGN.com in computer gaming) using micro norms. Clarke et al., (2023) devised a dual encoder framework (i.e., rule encoder and text encoder) incorporating meso norms for hate speech detection. Kumar et al. (2024) leveraged meso norms and large language models (e.g., GPT-3, Gemini Pro, LLaMA 2) for toxicity detection. Consequently, prior research has shown notable gaps: 1) there is scarce research on explicitly integrating community rules into these deep learning frameworks to facilitate content moderation and enhance the performance of content moderation; 2) most prior research is based on the rules from one single community or designed for a specific moderation task (e.g., hate speech, toxicity detection, and antisocial behavior detection), which lacks generalizability; 3) content moderation involving data collected from one single or individual communities confronts issues of imbalanced or biased data and instance scarcity; and 4) the definition of content inappropriateness, as determined by the regulations and rules of each specific community (Kotenko et al., 2017), lacks clarity and a foundational basis for regulating and make decisions about online content moderation. This deficiency undermines the transparency and interpretability of the deep network systems (Bunde, 2021).

To address the aforementioned research gaps, this research aims to answer the following questions by developing and empirically evaluating a novel content moderation model, namely *Community Rules-based Content Moderation-CRCM*:

1) How can we articulate community rules for designated online communities?

2) Can CRCM outperform other state-of-the-art methods in content moderation?

3) To what extent do the community rules improve the performance of content moderation?

This study makes two-fold contributions to content moderation research. First, we introduce a novel community rule-based content moderation framework that integrates the rules of online communities into deep learning models to enhance the performance of content moderation. There is little attention to community rules in developing content moderation models in the literature. Despite the potential of community rules, how to incorporate them into content moderation models remains a challenging issue. This research addresses the above challenge by proposing the framework and empirically demonstrating the superior performance of the framework. Second, unlike previous studies that concentrate on a single community, we extend our analysis to include the top ten most popular communities on Reddit. With this broader scope, we provide strong evidence for the model's generalizability across various online communities. These guidelines provide a solid foundation for the development of community rules, particularly addressing the cold-start setup challenges for new communities on social media platforms.

## 2. Related work

### 2.1. Content moderation in social media

Content moderation can be classified into three categories based on the phase of the intervention, including pre/proactive moderation, post moderation, and reactive moderation (L. Wang & Zhu, 2022).

Pre/proactive moderation, as elucidated by Habib et al., (2019), Schluger et al., (2022), and Taylor et al., (2020), centers on preemptively averting the presentation or circulation of contentious material. This approach entails a comprehensive evaluation of all content by both human moderators and automated systems before its publication or exposure to other users. It is primarily adopted in contexts characterized by heightened sensitivity or elevated risk, such as online platforms catering to children or forums focused on political discourse, which serves as a proactive measure to uphold desired standards of content quality and safety. In addition, post-moderation, as discussed by De Gregorio, (2020) Kamara et al., (2022), and Pan et al., (2022), entails the moderation of content subsequent to its publication. This approach may require less time investment compared to pre/proactive moderation and remains instrumental in mitigating the visibility of detrimental content across platforms. Furthermore, reactive moderation, as expounded upon by Alizadeh et al., (2022), Clune & McDaid, (2023), and Llansó, (2020), operates on the basis of evaluating content solely subsequent to its flagging or reporting by users. Under this framework, platforms rely on user-generated reports to initially identify potentially harmful content, subsequently taking appropriate actions if such content contravenes community standards or terms of service.

This research is focused on pre/proactive moderation by incorporating (predefined) community rules and user-generated post content to develop an automated content moderation framework to facilitate content moderation decision-making before user-generated post content becomes available to the public.

### 2.2. Deep learning-based content moderation approaches

Deep learning techniques, as evidenced by prior studies (e.g., Fu et al., 2023; Torfi et al., 2021), have emerged as exceptionally successful methods across various domains, opening a promising venue for managing online content (K. Wang et al., 2022, 2023). Deep learning-based methods for online content moderation span a wide spectrum of research topics, including abusive language recognition (Davidson et al., 2017; Founta et al., 2019), cyberbullying detection (Iwendi et al., 2023), and online misinformation identification (X. Zhou & Zafarani, 2020).

Deep learning-based approaches typically leverage hidden neural networks to process input features, ultimately making content moderation decisions for specific social media posts. More specifically, K. Wang et al., (2022) investigated the characteristics, degree, and efficiency of user engagements (i.e., users' comments) in moderating social media posts, which employed a pre-trained RoBERTa module for text embedding. In addition, Wang et al., (2023) developed a framework that incorporated the creditability and stance of users' comments into graph learning to facilitate content moderation. Furthermore, Park et al., (2022) used 97 subreddit classifiers, each with a four-layer network architecture to identify user comments that violate macro norms in comparison to the final decisions made by three independent, well-trained human annotators through a majority rule.

In the realm of content moderation, some other studies have designed task-oriented content moderation. For instance, Davidson et al., (2017) used TF-IDF-weighted n-gram features along with L2 regularization logistic regression for hate-speech detection. Founta et al., (2019) utilized a unified deep learning framework for detecting online abuse, separately processing a wide range of Twitter metadata to combine with hidden patterns within the text of tweets before developing a concatenation layer for classification. The text path employs a Recurrent Neural Network (RNN) with attention mechanisms, while the metadata path consists of six dense, fully connected neural network layers. In the study by Iwendi et al., (2023), four deep learning models were employed for content moderation, specifically targeting cyberbullying detection, with the BLSTM model achieving the best performance. Y. Zhou et al., (2020) fused the classification results of Embeddings from Language Models (i.e., 'ELMO') and Bidirectional Encoder Representations from Transformers (BERT), and Convolutional Neural Networks (CNN) for hate speech detection. Additionally, transformer-based methods (e.g., Tan et al., 2020; Wang et al., 2022) were explored to extract context clues from text information for moderation decision prediction.

Rule-based deep learning research on content moderation primarily focuses on a within-community moderation or a specific sub-task of moderation, as seen in studies by Cheng et al., (2015), Clarke et al., (2023), and Kumar et al., (2024), where only data from a single target community are involved or the model is only designed for a sub-task of content moderation (e.g., hate speech detection). Consequently, these approaches are effective within specific communities but face limitations of generalization when addressing cross-community or cross-subreddit content moderation. Compared with within-community moderation, cross-community moderation Chandrasekharan et al., (2019) is more robust in handling issues of data scarcity and imbalance.

## 3. A framework of community rules-based content moderation (CRCM)

### 3.1. Problem formulation

We formulate the content moderation problem as a binary classification task that classifies the user-generated content in social media as either being moderated or not. Let $c$ denote the textual content of a user-generated post, containing a post title and body. Given a list of rules $r$ specifically to a community, we focus on training a classification model augmented by community rules for content moderation decisions (see Equation 1):

$$y = f(c, r, \theta) \qquad (1)$$

where $y$ denotes a binary classification result of a target post $c$ augmented by community rules $r$, and $\theta$ denotes the set of parameters of the classifier $f(\cdot)$. More importantly, the proposed framework integrates the information of both community-specific rules and user-generated post content.

The overall framework of our proposed CRCM model is illustrated in Figure 1, which consists of three components: post and rule representation, classification, and rule-based affiliation measurement.

### 3.2. The CRCM framework

#### 3.2.1. Post representation

To represent the user-generated post content $c$, we leverage BERT (Devlin et al., 2019), a pre-trained embedding model that converts the text into vector representations $h_c$ in 768 dimensions. Specifically, we utilize the special classification token (i.e., $[CLS]$) added at the beginning of the word sequence in the BERT model, which usually serves as the aggregated sequence representation for an entire word sequence (Reimers & Gurevych, 2019; J. Wang et al., 2024).

### 3.2.2. Rule representation

Considering that some of the rules in the social media community are vague and may contain redundant content (e.g. 'Guidelines to prevent potential spoilers.' and 'Concerns spoiler guidelines'), we reorganize the rules by extracting topics from a list of rules in the domain.

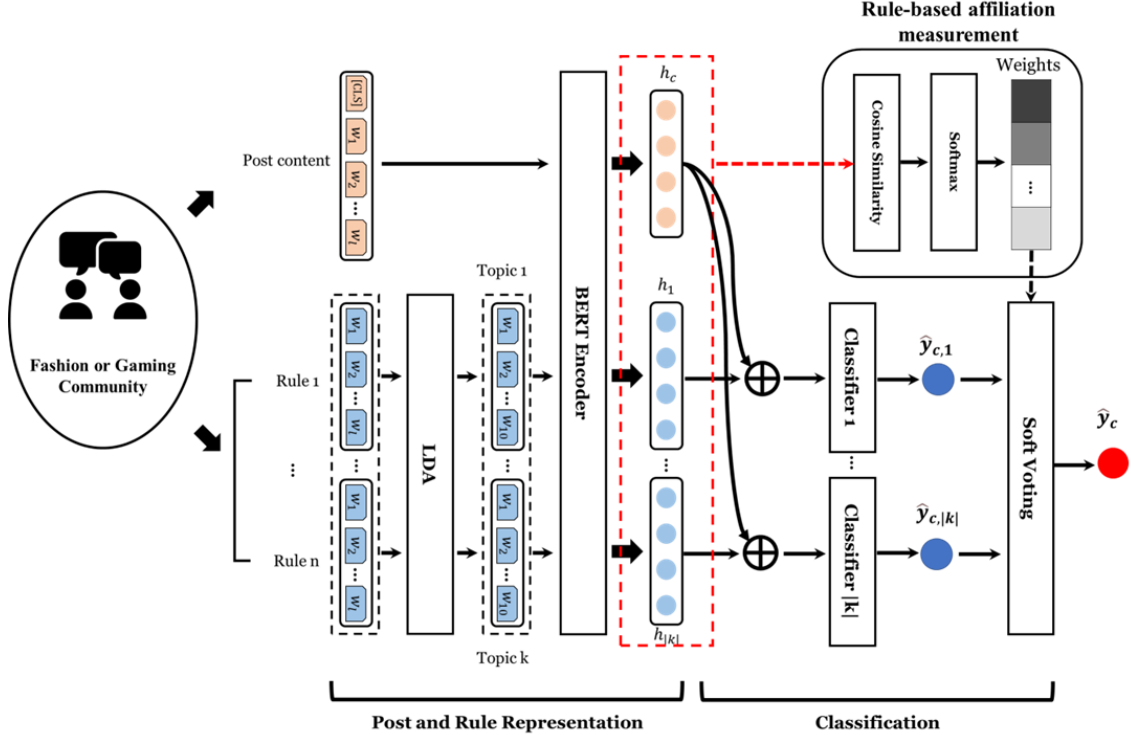

**Figure 1. The framework of Community Rules-based Content Moderation (CRCM)**

To be specific, we first combine all the rules collected from ten online communities within the same domain as a document and then extract topics from this document using the Latent Dirichlet Allocation (LDA) model (see Section 4.1). We depict a topic learned from the rule document with the probability distribution of words. To generate topic representations that concentrate on keywords in topics, we select the top ten words with the highest probability for each topic and aggregate their embeddings using Equation 2:

$$h_k = \sum_i p_i e_i \quad (2)$$

where $h_k$ denotes the representation of topic $k$, $e_i$ denotes the 768-dimensional BERT embedding vector of the top $i$th word with the highest probability in the topic, and $p_i$ denotes the normalized probability value of the top $i$th word. As a result, the rules can be represented by the combination of topic representations $r = [h_1 \dots h_{|k|}]$.

### 3.2.3 Classification

The classification for content moderation goes through a two-step process. We first design separate classifiers for each topic to generate an individual classification result and then use their ensemble as the final classification results via soft voting. To be specific, for each classifier, we concatenate the post representation $h_c$ with each topic representation of community rules $h_k$. In addition, a fully connected layer is designed to generate a probability value $\hat{y}_{c,k}$ of whether a target post $c$ augmented by rule topic $k$ should be moderated.

$$\hat{y}_{c,k} = \sigma(W_o[h_c; h_k] + b_o) \quad (3)$$

where $\sigma(\cdot)$ is the sigmoid activation function, $W_o \in \mathbb{R}^{1 \times 2d}$ is the learnable weight, $d$ is in 768-dimensionality and $b_o$ is the bias for the fully connected layer. We utilize the soft voting algorithm to aggregate the classification results from different rule topics with equal weights:

$$\hat{y}_c = \frac{\sum_k \hat{y}_{c,k}}{|k|} \quad (4)$$

We leverage the binary cross-entropy to optimize the binary classification model. The objective function of our model is shown below:

$$\mathcal{L} = \sum_{\hat{y}_c \in D} \hat{y}_c \log(\hat{y}_c) + (1 - \hat{y}_c)\log(1 - \hat{y}_c) + \mu\|\Delta\| \quad (5)$$

where $D$ denotes the set of training instances, and $\mu\|\Delta\|$ is parameter specific regulation hyper-parameters to prevent overfitting. By minimizing the loss value calculated in Equation 5, we tailor the model to accurately predict the probability of content moderation.

### 3.2.4. Rule-based affiliation measurement

One of the main limitations of applying soft voting is that the classifier treats all the rule topics equally, which is subject to the risk of introducing bias from less relevant rule topics. We argue that the classification model should prioritize the most relevant rule topics to the posts for the purpose of content moderation. For example, compared with the rule topic "No advertisement", the rule topic "To be polite" is more relevant and helpful for the classifier to determine whether a post with offensive content should be moderated. To address the above limitation, we propose a rule-based affiliation measurement method to adjust the voting weights across different rule topics. Specifically, we calculate the affiliation level between post content and rule topics via cosine similarity:

$$a_{c,k} = \frac{h_c \cdot h_k}{|h_c||h_k|} \quad (6)$$

Subsequently, we leverage a Softmax function to normalize the affiliation scores and adjust the voting weights of each rule.

$$w_{c,k} = Softamx(a_{c,k}) = \frac{\exp(a_{c,k})}{\sum_k \exp(a_{c,k})} \quad (7)$$

Accordingly, the final classification result can be calculated as the weighted sum of $\hat{y}_{c,k}$ based on the affiliation scores.

$$\hat{y}_c = \sum_k w_{c,k} \hat{y}_{c,k} \quad (8)$$

## 3.3. Evaluation

### 3.3.1. Data collection and pre-processing

We select Reddit as the primary source for data collection because it is one of the most prominent social media platforms facilitating social news aggregation, content evaluation, and interactive discussions (Jungherr et al., 2022). Subreddits, constituting distinct online communities, amalgamate a vast array of content within their specific domains. Consequently, our study focuses on the top 10 communities (i.e., subreddits) from each of the two thematic domains, with varying degrees of content moderation: fashion (35% moderated content) and gaming (13%).

Utilizing the Pushshift API[1], we conducted daily scraping of posts from 20 subreddits spanning the two diverse domains between August 24 and October 28, 2022, resulting in a total of 43,230 posts. As manual moderation entails significant time commitment, the effectiveness of content moderation is contingent upon various factors, including the nature and volume of content, the intricacy of moderation protocols, and the availability of human moderators. To augment the ecological validity of our study's findings, a subsequent round of data collection was performed using the PRAW API [2] two months later to ascertain the moderation status of the collected posts. The metadata of the collected data includes subreddit community rules, post content, post bodies, post time, etc.

To address the imbalance in the collected data due to the prevalence of legitimate content in Reddit posts, we employ the under-sampling approach. This procedure yielded a total of 5,958 posts, consisting of 1,152 from the fashion, and 4,806 from the gaming. Subsequently, all the gathered data within each domain were equally divided between moderate and non-moderate categories.

### 3.3.2. Baseline models

We selected the following four models for content moderation as baselines in this study due to their high relevancy to our study.

- Classifier97 (Park et al., 2022): The model comprises 97 neural network binary classifiers, each trained on data from one of the 97 online communities (i.e., subreddits). Each classifier features a four-layer neural network architecture, with an embedding layer and dense layers to make a binary prediction for content moderation.
- CB-BLSTM (Iwendi et al., 2023): This model employs a Bidirectional Long Short-Term Memory (BLSTM) architecture, which retains two separate input states—forward and backward—provided by two different LSTMs, to handle text sequences of posts.
- AbuDL (Founta et al., 2019): A Recurrent Neural Network (RNN) architecture is employed, specifically utilizing word-level RNNs for the raw text input. It is worth noting that the metadata was not taken into account in this study due to limitations in data availability.
- HATE-L2 (Davidson et al., 2017): This model leverages logistic regression with L2 regularization. The input features are pre-processed by lowercasing and stemming each post's content,

---

[1] https://github.com/pushshift/api

[2] https://praw.readthedocs.io/en/stable/

followed by creating TF-IDF-based n-gram representations.

### 3.3.3. Model evaluation

We employ a common set of evaluation metrics to assess model performances, including accuracy, precision, recall, and F1 measures. In addition, we implement 10-fold cross-validation, dividing the data into 10 folds with nine folds used for training and the remaining one for testing. The reported results are an average across these 10 folds.

We use the learning rate of 0.001; the dimension of BERT embedding is 768; and the dropout rate is 0.2 for all the models. All the models are optimized using the Adam optimizer.

## 4. Results

We report experiment results and additional analyses in this section.

### 4.1. Community rules

We present the community rules summarized from topic modeling outcomes alongside their coherence scores (i.e., a metric used to evaluate the quality of the topics generated by a topic model) in Figure 2. Notably, our results show that the topic modeling outcomes attain optimal coherence scores with six topics in the fashion domain and eight topics in the gaming domain.

Table 1 provides a summary of the community rules for both fashion and gaming domains, based on the highest coherence achieved.

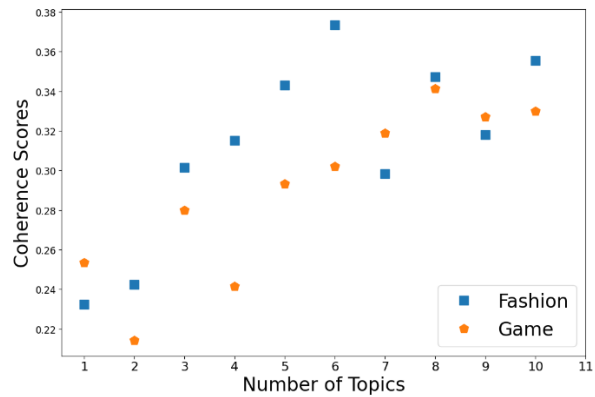

**Figure 2: Change of coherence score with the number of topics**

**Table 1. Two domain-specific community rules**

| | |
|---|---|
| *Fashion* | Details about permissions, content removal, and personal references. |
| | Posts comprising comments, courtesy, observation, characters, and memes. |
| | Photo rules and bans, with attention to messages and offenses. |
| | Tattoos-related posts focusing on deals, usage, and sales. |
| | Cosmetics-related inquiries, conversations, user engagements, and references to spam. |
| | Make-up-related posts inclusion requirements: products, content, and comments. |
| *Gaming* | Guidelines to prevent potential spoilers. |
| | Highly game-related posts. |
| | Discussions about the use of content within the community, including memes and discussions. |
| | The inclusion of specific game-related information in posts. |
| | Instructions on marking and handling spoiler content. |
| | Deals with rules for promotion, spam prevention, and engagement in the community. |
| | Discussions about trades, memes, and trolling. |
| | Instructions of potential bans. |

### 4.2. Model performance

We evaluate the effectiveness of our proposed content moderation model against four baseline models. Table 2 shows the model performances, and Table 3 presents the results of paired sample t-tests. The findings reveal that CRCM consistently surpasses all baseline models in terms of accuracy and precision, at a significance level of .001. Moreover, CRCM consistently achieves superior recall compared to all baseline models at a minimum significance level of .05, except for the comparison to the AbuDL model ($p>.1$) in the gaming domain. Furthermore, CRCM achieves a higher F1 score than all baseline models, with at least a .01 significance level.

### 4.3. Ablation study

To investigate the effects of soft voting weight (i.e., affiliation) and community rules on model performance, we conducted an ablation study by systematically removing each component from the CRCM model separately and performing multiple paired sample t-tests. Figure 3 plots the model performances for the fashion and gaming domains respectively, and Table 4 details the paired t-test results, including mean differences and t-statistics. The findings indicate that the CRCM model achieves the optimal performance when both affiliation and community rules are incorporated. Both components significantly enhance model performance at a significance level of .001, with community rules contributing the most substantial improvement.

**Table 2. Model performances of the CRCM and the baseline models**

| Domains | Models | Accuracy | Precision | Recall | F1 |
|---|---|---|---|---|---|
| *Fashion* | Classifier97 | 49.99% | 49.84% | 52.43% | 50.94% |
|  | CB-BLSTM | 61.90% | 61.53% | 64.08% | 62.68% |
|  | AbuDL | 51.64% | 51.07% | 59.25% | 54.14% |
|  | HATE-L2 | 63.12% | 62.92% | 64.07% | 63.35% |
|  | ***CRCM*** | **69.19%** | **69.12%** | **69.61%** | **69.23%** |
| *Gaming* | Classifier97 | 50.79% | 50.72% | 51.77% | 51.06% |
|  | CB-BLSTM | 56.03% | 56.08% | 57.87% | 56.26% |
|  | AbuDL | 49.92% | 49.34% | 62.35% | 51.92% |
|  | HATE-L2 | 63.25% | 63.13% | 63.75% | 63.42% |
|  | ***CRCM*** | **69.91%** | **69.41%** | **71.28%** | **70.31%** |

Note: the best results are highlighted in boldface.

**Table 3. Performance comparisons between the CRCM and the baseline models**

| Domains | Models | Accuracy | Precision | Recall | F1 |
|---|---|---|---|---|---|
| *Fashion* | Classifier97 | -10.194*** | -8.662*** | -6.185*** | -8.200*** |
|  | CB-BLSTM | -8.169*** | -6.940*** | -2.293* | -6.274*** |
|  | AbuDL | -10.732*** | -14.030*** | -1.947* | -4.967*** |
|  | HATE-L2 | -6.698*** | -6.097*** | -4.851*** | -6.300*** |
| *Gaming* | Classifier97 | -22.968*** | -21.170*** | -9.066*** | -13.803*** |
|  | CB-BLSTM | -20.299*** | -15.783*** | -3.448** | -7.774*** |
|  | AbuDL | -27.391*** | -24.616*** | -1.024 | -3.688** |
|  | HATE-L2 | -8.047*** | -7.056*** | -9.701*** | -8.647*** |

t-statistic; ***: $p<.001$; **: $p<.01$; *: $p<.05$.

**Table 4. Performance comparisons of the CRCM model and its ablated models**

| Domains | Models | Accuracy | Precision | Recall | F1 |
|---|---|---|---|---|---|
| *Fashion* | Affiliation | -.065*** | -.060*** | -.080*** | -.071*** |
|  | Community Rules | -.081*** | -.078*** | -.078*** | -.081*** |
| *Gaming* | Affiliation | -.063** | -.045*** | -.111*** | -.082*** |
|  | Community Rules | -.145*** | -.142*** | -.121*** | -.136*** |

Note: when community rules are ablated, it refers to the CRCM model with the community rules component removed. It is the same for affiliation. ***: $p<.001$.

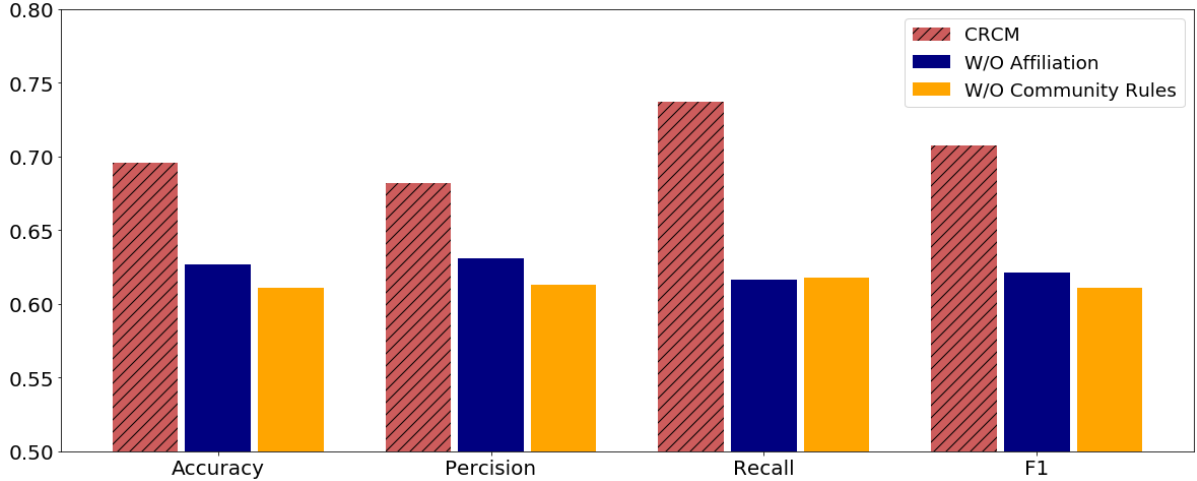

(a) Fashion

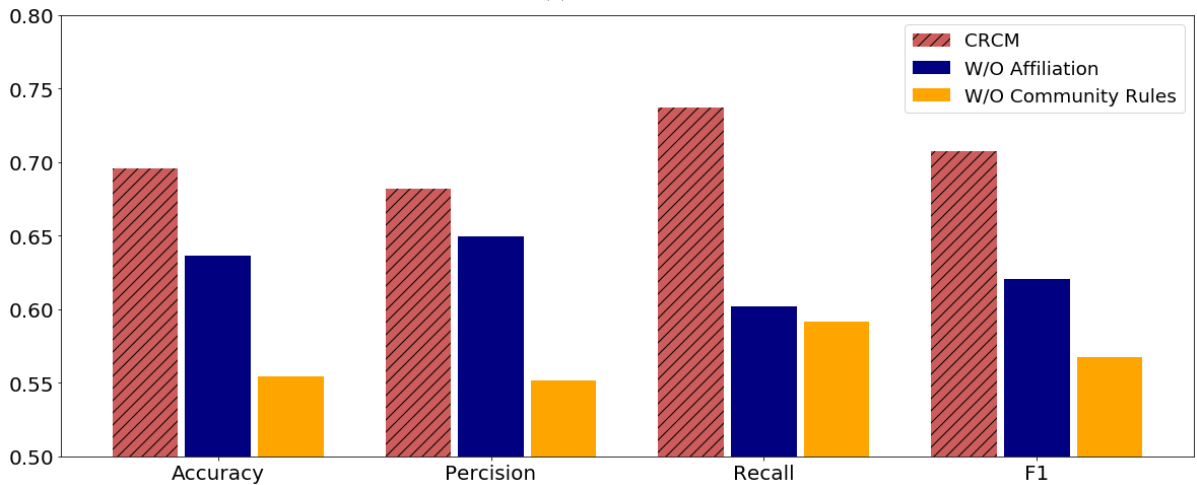

(b) Gaming

**Figure 3. Ablation Experiment Results**

## 5. Discussion

This study underscores the significant benefits of incorporating online community rules into deep learning models for content moderation, thereby fostering healthier, safer, and more sustainable online environments. By integrating with domain-specific rules from various online communities, our proposed CRCM framework significantly outperforms the baseline models across all evaluation metrics and two different domains. In addition, the results of our ablation study also reveal that afflated community rules greatly contribute to performance improvement in content moderation. The findings of this study suggest that the development of a comprehensive set of community guidelines not only helps improve the accuracy and fairness of content moderation but also lays a solid foundation for new online communities to create rules or norms, and for established communities to refine their rules, ultimately enhancing user engagement and trust and reputations of online communities on social media platforms.

This research can be extended in multiple directions in future research. First, we focus on pre/proactive moderation – the most challenging type of moderation task in this study. Post moderation and reactive moderation are also expected to benefit from exploiting community rules, which need to be validated in future research. Second, it is worth exploring the potential of large language models (e.g., GPT, Gemini, and LLaMA) in content moderation tasks. Third, future research should explore hybrid systems that seamlessly team up human moderators and AI models to further enhance model performance.

Lastly, community rules evolve over time, and consequently, the content moderation models should continuously adapt themselves to community norms over time.

## Acknowledgement

This work was partially supported by Truist Research Grant and UNC Charlotte Graduate School Summer Fellowship. Any opinions, findings, or recommendations expressed here are those of the authors and are not necessarily those of the sponsors of this research.